\documentclass[a4paper,twocolumn,english,pre,aps,floatfix,showpacs]{revtex4}
\usepackage[T1]{fontenc}
\usepackage[latin1]{inputenc}
\usepackage{amsmath}
\usepackage{graphicx}
\usepackage{amssymb}

\makeatletter
\usepackage{graphics}
\usepackage{amssymb}
\usepackage[latin1]{inputenc}
\usepackage{latexsym}
\usepackage{natbib}
\usepackage{verbatim}
\usepackage{ae,aecompl}
\def\kbar{{\mathchar'26\mkern-9mu k}}

\usepackage{babel}
\makeatother
\begin{document}

\title{Phase-space reconstruction of an atomic chaotic system}

\author{Hugo L. D. de Souza Cavalcante}

\author{Carlos Renato de Carvalho}
\altaffiliation[Permanent address: ]{Instituto de Física, Universidade Federal 
do Rio de Janeiro, Caixa Postal 68528, 21941-972 Rio de Janeiro, RJ, Brazil} 
\email{crenato@if.ufrj.br. }

\author{Jean Claude Garreau}

\affiliation{Laboratoire de Physique de Lasers, Atomes et Mol{\'e}cules, Universit{\'e}
des Sciences et Technologies de Lille, Bat. P5, F-59650 Villeneuve
d'Ascq Cedex, France}

\homepage{http://www.phlam.univ-lille1.fr/atfr/cq}

\date{19 March 2005}%\today \, \currenttime}
\begin{abstract}
We consider the dynamics of a single atom submitted to periodic pulses
of a far-detuned standing wave generated by a high-finesse optical
cavity, which is an atomic version of the well-known ``kicked rotor''.
We show that the classical phase-space map can be ``reconstructed''
by monitoring the transmission of the cavity. We also studied the
effect of spontaneous emission on the reconstruction, and put limits
to the maximum acceptable spontaneous emission rate.
\end{abstract}


\maketitle

\section{Introduction
\label{sec:Introduction}}

The kicked rotor (KR) is a widely explored system that
has a paradigmatic status in studies of classical 
\cite{Chirikov_ChaosClassKR_PhysRep79}
and quantum chaos \cite{Casati_LocDyn_79,Izrailev_LocDyn_PREP90}.
In recent years, the advent of laser cooling of atoms allowed the
realization of an atomic version of the KR 
\cite{Zoller_LDynTheo_PRA92,Raizen_LDynNoise_PRL98},
which has been used in experiments by many groups 
\cite{Raizen_LDynNoise_PRL98,Christ_LDynNoise_PRL98,DArcy_AccModes_PRL99,AP_Bicolor_PRL00}.
This consists simply in placing laser-cooled atoms in a pulsed laser
standing wave. In adequate (but rather general) conditions, the atomic
center of mass motion ``feels'' the presence of the standing wave
as an ``optical potential'' which varies sinusoidally in space.
One can then easily modulate (with period $T)$ the light intensity
in the form of short pulses to obtain a hamiltonian of the form %
\footnote{For a detailed description of a typical experimental setup see e.g.
\cite{AP_ChaosQTransp_CNSNS_2003}.%
}
\begin{equation}
H=\frac{p^{2}}{2M}+V_{0}\cos(2k_{L}x)\sum_{n=0}^{N-1}\delta_{\tau}(t^{\prime}-nT),
\end{equation}
where $M$ and $p$ are resp. the atom mass and center of mass momentum, 
$V_{0}$ is the amplitude of the optical potential
(which is proportional to $I/\Delta$, where $I$ is the intensity
of the standing wave and $\Delta$ the laser-atom detuning), $k_{L}=2\pi/\lambda_{L}$
the wave number of the standing wave (in the $x-$direction), and
$\delta_{\tau}$ a square function of duration $\tau$ and amplitude
$1$. In the limit $\tau\rightarrow0$, or, more exactly, in the limit
\begin{equation}
\frac{\left\langle p^{2}\right\rangle ^{1/2}\tau}{M}\ll\lambda_{L},
\label{eq:KRValidity}
\end{equation}
one can put $\delta_{\tau}/\tau\rightarrow\delta$ and one obtains
exactly the hamiltonian of the KR. In order to obtain this hamiltonian
in its standard form, we perform the normalizations $t^{\prime}/T\rightarrow t$,
$2k_{L}x\rightarrow X$, $(2k_{L}T/M)p\rightarrow P$:
\begin{equation}
H=\frac{P^{2}}{2}+K\cos X\sum_{n=0}^{N-1}\delta(t-n)
\label{eq:KRHamiltonian}
\end{equation}
where the only free parameter is the normalized kick strength $K=4V_{0}k_{L}^{2}T\tau/M$.

Many of the recent experimental studies were aimed at \emph{quantum}
aspects of the KR dynamics, as ``dynamical localization'' 
\cite{Raizen_LDynNoise_PRL98,AP_SubFourier_PRL02},
quantum resonances \cite{Darcy_QRes_PRL01}, or ``chaos-assisted''
tunneling \cite{Raizen_ChaosAssistTunnel_Science01,HRDunlop_ChaosAssitTunnel_Nature01}.
Quantum effects in such system manifest for times larger than the ``Ehrenfest time",
when quantum dynamics begins to take over classical dynamics. This
is the domain of ``quantum chaos'', which is defined as the \emph{quantum}
dynamics of systems which are chaotic in the classical limit. It is
well known that the quantum behavior of such systems bears little
relation to the classical chaotic behavior of their classical counterparts,
basically because Schr{\"o}dinger equation is linear. One of the most
exciting questions about quantum chaos is ``How the initially classical
dynamics and latter quantum dynamics diverge from one another?''
Clearly, the critical point in this divergence is situated between
the Ehrenfest time $t_{E}\sim T\ln(1/\kbar)$ and the so-called ``localization
time" $t_{L}\sim T(K/\kbar)^{2}$, where $\kbar$ is the effective
Planck constant %
\footnote{In the usual version of the atomic KR $\kbar=4\hbar k_{L}^{2}T/M$;
it can thus be controlled by changing the pulse period $T$.%
}. Fortunately, in the atomic KR the parameters can be widely changed
which allows one to adjust these times to experimentally convenient
scales. However, even if the adequate time domain is accessible, it
is still difficult to put into evidence the transition between classical
and quantum regimes because, usually, the kind of measurements performed
on a microscopic system is different from -- and often not simply
related to -- the kind of measurement performed on macroscopic systems.
Typically, studies of classical dynamics rely on the notion of trajectory
(phase-space maps, first-return maps, Poincar{\'e} sections...) whereas
studies of quantum dynamics emphasize mean quantities and probability
distributions. The purpose of the present paper is to describe and
study an experimentally realizable situation in which reconstruction
of the atom center of mass dynamics can be done, thus reconciliating
quantum- and classical-type measurements. Hopefully, such a setup would
provide a better understanding of the above-mentioned transition between
classical and quantum dynamics. Throughout this paper, we shall treat
the atomic motion as a classical motion: we shall consider
that the atomic center of mass position and momentum are enough to
characterize the dynamics. This is true as long as the dynamics is
essentially classical, that is, for times shorter than the localization
time, or as far as the sources of decoherence -- mainly spontaneous
emission in the present case -- act often enough to provoke a quantum
coherence collapse before interference effects can develop and appreciably
change the classical behavior.

\section{Kicked rotor dynamics}

\begin{figure} % Fig 1
\begin{center}\includegraphics[%
  width=8cm,
  keepaspectratio]{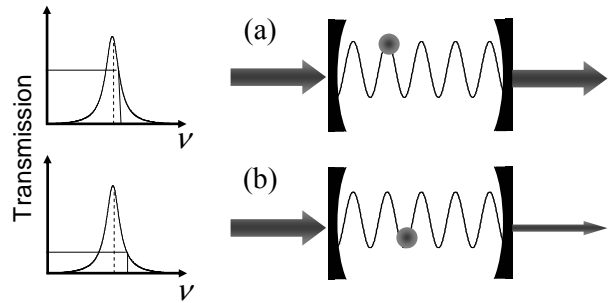}\end{center}
\caption{
\label{fig:schematic}Schematic view of the proposed setup. If the
atom is at a crest (a) or at a node (b) of the cavity electromagnetic
field, the Kerr nonlinear refractive index produced is different,
the resonance condition of the cavity is modified, and the transmitted
intensity changes, allowing detection of the longitudinal position
of the atom.}
\end{figure}

In recent years, the quality of optical cavities has been dramatically
improved. It is now possible to build cavities of very
high finesse \cite{Kimble_AtomMotion_S00,Rempe_AtomMotion_PRL99}.
In such a cavity, the index of refraction generated by {\em one} atom
is enough to change mesurably the resonance condition, and the cavity
can thus be used to ``count atoms one by one'' \cite{Kimble_CavityCountAtom_PRL04}.
As an atom moves inside the cavity, it ``probes'' regions where the
radiation intensity is not the same. For example, cavity modes usually
have a gaussian profile, whose intensity is larger at the center than
in the border. In the same way, as the field inside the cavity forms a
standing wave, changing the longitudinal position of the atom also
changes the intensity to which it is submitted (Fig.~\ref{fig:schematic}). 
This is the situation we shall consider in the present paper.
Because of the Kerr effect, the refractive index of
the atom is modified by the intensity of the radiation to which it
is exposed, and the changing of the refractive index in turn changes
the cavity resonance condition. By
measuring the cavity transmission one can extract information on the
position of the atom. This is an ``autoconsistency'' problem:
the change in the resonance condition changes the cavity field, that
in turn changes the atom refractive index. This means that the amplitude
of the kicks felt by the atom will also depend on its position, which
is in contradiction with the standard definition of the KR. We shall
take this fact into account in our simulations, but will arrange parameters
in order to make the variation of the kick strength $K$ small enough
to not disturb too much the dynamics of the KR.

An experimental limitation that must also be taken into account is
that one cannot change instantaneously the radiation intensity inside
a cavity. The ``lifetime'' of a photon inside the cavity is roughly
$\tau_{c}=L/(c\mathcal{T})$, where $\mathcal{T}$ is the (intensity)
transmission coefficient of the mirrors and $L$ the length of the
cavity. The duration $\tau$ of the kicks shall thus be at least of
a few $\tau_{c}$. However, the validity condition for identifying
the kicks with a delta function, Eq.~(\ref{eq:KRValidity}), must
also be satisfied. This implies that
\begin{equation}
\frac{2hL}{\lambda_{L}^{2}Mc}\sqrt{N}\ll\mathcal{T}
\end{equation}
where we used the order of magnitude value $\left\langle p^{2}\right
\rangle ^{1/2}\sim2\hbar k_{L}\sqrt{N}$
after $N$ kicks are performed 
\footnote{This order of magnitude value is obtained by considering the motion
as perfectly diffusive in the momentum space (which is characteristic
of a chaotic behavior): the spreading in $p$ increases as the
square-root of the number of ``steps'' (kicks) $\sqrt{N}$, each
step corresponding to a momentum shift equal to the exchange of a
photon between the two counterpropagating waves forming the standing
wave $2\hbar k_{L}$.%
}. Thus given $\mathcal{T}$, there is a maximum number of kicks (i.e.
a maximum duration of the experiment), 
$N_{max}\sim\left[\lambda_{L}^{2}Mc\mathcal{T}/(2hL)\right]^{2}$.
For cesium this gives $N_{max}\sim(\mathcal{T}/L)^{2}\times 10^{9}$ m$^{2}$.
This condition is not hard to satisfy: for $\mathcal{T}=10^{-6}$
and $L=10^{-4}$ m, $N_{max}\sim10^{5},$ whereas experiments currently
last only for a few hundred kicks.

Integrating the equations of motion obtained from Eq.~(\ref{eq:KRHamiltonian})
over one period $T=1$ produces the so-called ``standard map'' 
\cite{Chirikov_ChaosClassKR_PhysRep79}
\begin{subequations}
\begin{equation}
X_{n+1}=X_{n}+P_{n}
\label{eq:KReqX}
\end{equation}
\begin{equation}
P_{n+1}=P_{n}-K_{n+1}\sin X_{n+1}.
\label{eq:KReqP}
\end{equation}
\label{eq:KReqs}
\end{subequations} where we took into account the
fact that, due to the nonlinear Kerr effect, $K_{n}=K(X_{n})$: the
intensity of the kick \emph{depends} on the position of the atom.
In constructing the phase-space portrait of the
kicked rotor, it is usual to simplify the display by taking all the
quantities ``modulus $2\pi"$ (we shall represent the modulus $2\pi$
of a given quantity $A$ by $A[2\pi]$). This is due to the fact taht
the potential is periodic in space, of period $2\pi$ in reduced
units. This means that if $P_{n}=2\pi q+p$, with $p<2\pi$ and $q$
integer, then, after Eq.~(\ref{eq:KReqX}), $X_{n+1}=X_{n}+p+2\pi q$. 
Thus $X_{n+1}[2\pi]=X_{n}[2\pi]+p$: the physics is the same for
momenta differing by an integer multiple of $2\pi.$ We shall thus,
according to the usual convention, draw phase-space maps by plotting
$X[2\pi]$ \emph{vs.} $P[2\pi]$. We show a typical plot of the KR
phase-space in Fig.~\ref{fig:KRphasemap}(a).

\section{Cavity dynamics}

A schematic view of the proposed experiment is shown in Fig.~\ref{fig:schematic}.
The atom is placed inside a high-finesse cavity, and its presence
shifts the resonance condition of the cavity in a nonlinear way: this
shift depends on the radiation intensity seen by the atom, and thus
on its longitudinal position. The intensity transmitted through the
cavity can thus be related to the position of the atom. However, this position
is defined only with respect to the closest node; shifting the position
of the atom by a period of the standing wave produces the same
transmission signal
\footnote{We suppose that the atom completely relaxes between kicks
so that there are no memory effects. This is justified by the fact
that typical kick periods are $T~20 ms$ that are much larger than
any atomic time-scale.}.

If a laser beam of wavenumber $k_{L}$ intensity $I_{0}$ is injected
into an empty Fabry-Perot cavity made of two identical mirrors a distance
$L/2$ apart, of transmission coefficient $\mathcal{T}$ and reflection
coefficient $\mathcal{R}=1-\mathcal{T}$, the intracavity intensity
is given by the well-known Airy function
\begin{equation}
I_{c}=\frac{1/\mathcal{T}}{1+F^{2}\sin^{2}\left(\frac{k_{L}L_{opt}}{2}\right)}I_{0},
\label{eq:Icavity}
\end{equation}
 leading to a resonance Rabi frequency
\begin{eqnarray}
\Omega^{2}&=&\Omega_{0}^{2}\cos^{2}\left(k_{L}x\right)\\
&=&\left(\frac{3\pi\Gamma}{\hbar ck_{L}^{3}}\right)\frac{1/\mathcal{T}}{1+F^{2}\sin^{2}
\left(\frac{k_{L}L_{opt}}{2}\right)}I_{0}\cos^{2}\left(k_{L}x\right)
\label{eq:RabiAiry}
\end{eqnarray}
where $F=2\sqrt{\mathcal{R}}/\mathcal{T}$ is the finesse of the cavity
divided by $\pi/2$ and $\Gamma$ the natural width of the atomic
transition. The parameter $K$ [Eq.~(\ref{eq:KRHamiltonian}] turns out to be 
\cite{Raizen_LDynNoise_PRL98}:
\begin{eqnarray*}
K & = & \frac{1}{2}\omega_{r}T^{2}\frac{\Omega_{0}^{2}}{\Delta}\\
 & = & \left(\omega_{r}T\right)\left(\Gamma T\right)
\left(\frac{\Gamma}{\Delta}\right)\frac{1/\mathcal{T}}{1+F^{2}\sin^{2}
\left(\frac{k_{L}L_{opt}}{2}\right)}\frac{I_{0}}{I_{s}}
\end{eqnarray*}
where $I_{s}\equiv\hbar ck_{L}^{3}\Gamma/(6\pi)$ is the saturation
intensity $\sim2$ mW/cm$^{2}$ for cesium, $\omega_{r}=\hbar k_{L}^{2}/(2M)$
is the recoil frequency ($\omega_{r}/(2\pi)\sim2$ kHz for cesium)
and $\Delta$ the laser-atom detuning.

In Eq.~(\ref{eq:RabiAiry}), $L_{opt}=L+\Delta L$ is the optical
length of the cavity, $L$ is its physical length and $\Delta L$
is the correction due to the presence of the atom:

\begin{eqnarray*}
k_{L}\Delta L & = & \frac{3\pi\Gamma}{2k_{L}^{2}L_{at}^{2}}
\frac{\Delta}{\Delta^{2}+\frac{\Omega^{2}}{2}+\frac{\Gamma^{2}}{4}}\\
 & = & \left(\frac{3\pi}{2k_{L}^{2}L_{at}^{2}}\right)
\frac{\Delta/\Gamma}{\frac{\Delta^{2}}{\Gamma^{2}}+
\frac{\Omega^{2}}{2\Gamma^{2}}+\frac{1}{4}}
\end{eqnarray*}
where $L_{at}$ is the linear size of the effective volume occupied
by the atom. The definition of this volume depends on the details
of the experiment being performed. In the case of laser-cooled atoms,
we can take this volume as the volume of the atom cloud in a magneto-optical
trap, roughly $L_{at}\sim100$ $\mu$m, which gives 
$3\pi/[2(k_{L}L_{at})^{2}]\sim10^{-5}.$
For $\Delta\approx10^{3}\Gamma$, which is a typical value, the dephasing
$k_{L}\Delta L$ induced by a lone atom is roughly 10$^{-8}$, which
is hard to detect directly. The effect is however enhanced by the
presence of the cavity.

The detected signal is the intensity transmitted through the cavity,
given by
\begin{equation}
S=\frac{1}{1+F^{2}\sin^{2}\left(\frac{k_{L}L_{opt}}{2}\right)}I_{0}=I_{c}/\mathcal{T}
\end{equation}
which depends on the radiation intensity seen by the atom through
$L_{opt}$ , and thus on the atom position. If $F\gg1$, the half-width
of the Airy peak corresponds roughly to
\begin{equation}
F^{2}\sin^{2}\left(\frac{k_{L}L_{opt}}{2}\right)\approx F^{2}
\frac{\left(k_{L}L_{opt}\right)^{2}}{4}=1,
\end{equation}
thus
\begin{equation}
k_{L}L_{opt}=\frac{2}{F}.
\end{equation}
One easily finds
\begin{eqnarray*}
\left|\frac{\Delta I_{t}}{I_{0}}\right| & \approx & \frac{F^{2}k_{L}L}{4}
\left|k_{L}\Delta L\right|=\frac{F}{2}\left|k_{L}\Delta L\right|
\end{eqnarray*}
where we supposed $k_{L}L\approx2/F$ . Then, taking $k_{L}\Delta L\approx10^{-8}$,
an experimentally realistic value \cite{Kimble_AtomMotion_S00} of $F\approx10^{6}$
produces a detectable 1\%-variation of the transmitted intensity %
\footnote{A larger value of the transmission variation is not desirable, as
it implies a larger spatial variation of $K$.%
}.

Techniques based on the Kerr nonlinearity similar to that proposed
here have been recently used for monitoring the \emph{radial} motion
of an atom trapped in high-finesse cavity 
\cite{Kimble_AtomMotion_S00,Rempe_AtomMotion_PRL99}.
In our case, it is the \emph{longitudinal} motion of the atom that
should be monitored. This is complicated by the periodic character
of the standing wave inside the cavity. We shall discuss
in the next section the algorithm we developed to deal with this problem.

\section{The reconstruction algorithm}

\begin{figure} % Fig. 2
\begin{center}\includegraphics[%
  width=4cm,
  keepaspectratio]{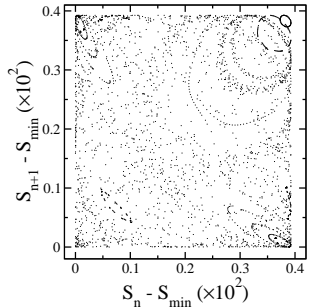}
\end{center}
\caption{
\label{fig:Transmission}First return map $S_{n}\times S_{n+1}$
for the cavity-transmission signal, with $F=10^{6}$. Structures
are clearly visible, but no direct relation to the topology
of the phase-space of the KR (Fig.~\ref{fig:KRphasemap}a) can be drawn. 
}
\end{figure}

\begin{figure} % Fig. 3
\begin{center}\includegraphics[%
  clip,
  width=8cm,
  keepaspectratio]{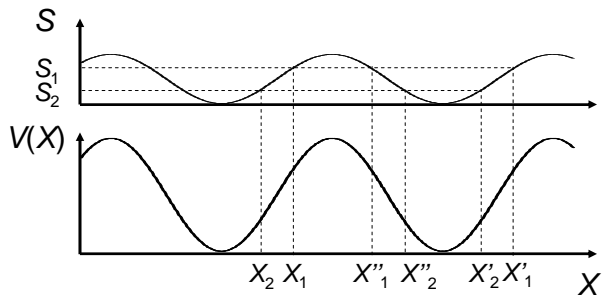}\end{center}
\caption{
\label{Fig:PositionTransmission}Correspondence between the transmission
signal and the atom longitudinal position inside the cavity.}
\end{figure}

\begin{figure} % Fig. 4
\begin{center}\includegraphics[%
  width=7.5cm]{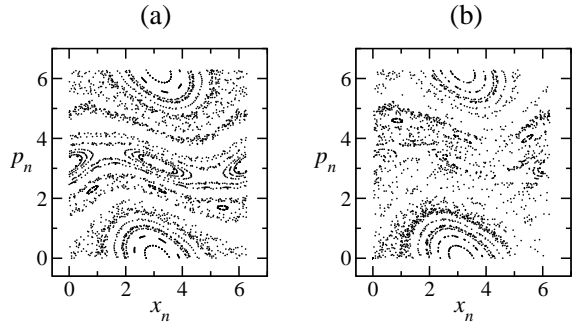}\end{center}
\caption{
\label{fig:KRphasemap}Comparison between the original (a) and the
reconstructed (b) KR phase space-map. Parameters are $K=0.8$, $F=10^{6}$.
Larger structures are easier to reconstruct than smaller ones.
}
\end{figure}

In order to reconstruct the phase-space map from the cavity transmission
signal, we suppose that the transmission value corresponding to each
kick $n$ is recorded, forming a temporal series $S_{n}$. The information
given by the bare signal is however incomplete. Fig.~\ref{fig:Transmission}
displays the first return map corresponding to the temporal series
$S_{n}$. Its topology do not evoke at all the KR phase-space map
shown in Fig.~\ref{fig:KRphasemap}a. Let us consider the information
that can be extracted from two successive values of the cavity transmission,
$S_{1}$ at kick $1$ and $S_{2}$ at kick 2. Referring to 
Fig.~\ref{Fig:PositionTransmission},
we see that the cavity transmission leaves an indeterminacy on the
position, as the points $X_{1}$, $X_{1}^{\prime}$ and $X_{1}^{\prime\prime}$
correspond to the same transmission. Generally two classes of points
lead to the same transmission
\begin{subequations}
\begin{equation}
X\quad\mathrm{and}\quad X^{\prime}=X+2\pi q
\label{eq:UndetX1}
\end{equation}
\begin{equation}
X\quad\mathrm{and}\quad X^{\prime\prime}=2\pi q-X.
\label{eq:UndetX2}
\end{equation}
\label{eq:XUndets}
\end{subequations}
($q$ is an integer).

\begin{widetext}
\begin{center}
\begin{table}[hbt]
%\begin{center}
\begin{ruledtabular}
\begin{tabular}{cccc}

\hline

$X$ & $ X_1 $ & $X_1+2\pi q_1$ & $ 2\pi q_1 - X_1 $ \\

\hline 

$X_2$ & $X_2-X_1 $ & $ X_2-X_1-2\pi q_1 $ & $X_2+X_1-2\pi q_1$ \\

$X_2+2\pi q_2 $ & $X_2-X_1 +2\pi q_2 $ & $ X_2-X_1+2\pi (q_2-q_1) $ & $X_2+X_1+2\pi (q_2-q_1)$ \\

$2\pi q_2-X_2$ & $-(X_2+X_1) +2\pi q_2 $ & $ -(X_2+X_1)+2\pi (q_2-q_1) $ & $-(X_2-X_1)+2\pi (q_2-q_1)$ \\

\hline
\end{tabular}
\end{ruledtabular}
%\end{center}
\caption{
\label{tab:xp}Possible momentum ($X_{2}-X_{1})$ determinations
for different choices of the position determination ($q_{1}$ and
$q_{2}$ are integers). Eliminating the terms of the form $2\pi\times$(integer
number), the only possibilities are $x_{2}+x_{1}$, $-(x_{2}+x_{1})$,
$x_{2}-x_{1}$ and $-(x_{2}-x_{1})$, with $x_{i}=X_{i}[2\pi]$. Considering
that trajectories corresponding to momenta $P$ and $-P$ are identical
in the phase-space map leaves only two choices, $x_{2}+x_{1}$ and
$x_{2}-x_{1}$.}
\end{table}
\end{center}
\end{widetext}

However, as it is $X[2\pi]$ that is displayed in the phase-space
map, the value of the integer number $q$ in Eqs. (\ref{eq:XUndets})
is irrelevant. This leaves us with just two possible values, $x$
and $2\pi-x$, where $x=X[2\pi]$. As the motion of the atom is free
between two kicks, the momentum associated to the kick $n$ is $P_{n}=X_{n+1}-X_{n}$.
Moreover, we note that $P$ and $-P$ lead to the same trajectory
in the phase portrait. Introducing the relevant (for phase-portrait-plotting
purposes) value $p_{n}=P_{n}[2\pi]$, we find that there are only
two distinct choices for the momentum: $p_{1}=x_{2}-x_{1}$ or $p_{1}=x_{2}+x_{1}$
(see table \ref{tab:xp}). We must thus chose arbitrarily one of the
two. If we now include the next value $x_{3}$ of the position (with
its own arbitrary choice of the determination), we can deduce in the
same way a value for $p_{2}$. But Eq.~(\ref{eq:KReqP}) implies that
the successive values $p_{1}$ and $p_{2}$ of the momentum must satisfy
\begin{equation}
p_{2}=p_{1}-\left(K_{2}\sin x_{2}\right)[2\pi],
\label{eq:KReqPMod2pi}
\end{equation}
which allows us to test the coherence of the choices.

In practice, our algorithm proceeds as follows: \emph{i}) For each
point $S_{n}$ in the time series of cavity transmission, we first
estimate a value for the intracavity intensity $I_{c}^{(0)}$ using
Eq.~(\ref{eq:Icavity}) with $L_{opt}=L$. This value of $I_{c}^{(0)}$is
then used to estimate the atom position $x_{n}^{(0)}$ and to calculate
a first correction to $L$, $L^{(1)}=L+\Delta L_{1}$, which in turn
is used to determine a corrected value $I_{c}^{(1)}$ and a corrected
value of the position $x_{n}^{(1)}$. The process is iterated until
it converges to produce the value of the position corresponding to
the $n^{th}$ kick, $x_{n}$. Ten steps are usually enough to insure
the convergence of the process. \emph{ii}) We choose arbitrarily between
the two possible determinations $x_{n}$ or $2\pi-x_{n}$ of the position
and calculate the corresponding value of $p_{n}$ (choosing arbitrarily
between $x_{n}-x_{n-1}$ and $x_{n}+x_{n-1}$). \emph{iii}) Each time
three successive momentum values have been determined, we use Eq.
(\ref{eq:KReqPMod2pi}) to test the coherence of the choices. If this
condition is violated, we recalculate the last three points with a
different choice of the position determinations. \emph{iv}) Once
a trajectory is reconstructed, one starts again with another time
series corresponding to another set of initial conditions.

Fig.~\ref{fig:KRphasemap} compares the exact KR phase-space map with
the reconstructed one. We see that the agreement is good. The
number of restarts due to the coherence test is 380 over 3000 data
points.

\section{Effect of the spontaneous emission}
The good reconstruction obtained in the preceding section is
however impossible to achieve in a real experiment, because of the
unavoidable presence of spontaneous emission in the system. However, the
spontaneous emission rate can be controlled -- to a certain extent -- in such systems, 
because it roughly scales with the inverse of the \emph{square} of the laser-atom detuning,
whereas the optical potential amplitude $V_{0}$ scales with the inverse
of the detuning; one can thus adjust the parameter in such way that
spontaneous emission level is tolerable during the experiment duration.
In the one dimensional situation we are considering, the effect of
spontaneous emission is to add to the value of $p$ in the average
a unit of photon momentum $\pm\hbar k_{L}$ (corresponding to 
$\kbar/2=2 \hbar k_L^2 T/M$ in reduced units), with an arbitrary sign,
each time it happens. We introduced
spontaneous emission in our simulations by a simplified Monte-Carlo
procedure. For each pulse of duration $\tau$, we calculate the 
probability for a fluorescence cycle to happen
\begin{equation}
g_{a}=\frac{\Gamma}{2}\frac{\Omega^{2}/2}{\Delta^{2}+
\frac{\Omega^{2}}{2}+\frac{\Gamma^{2}}{4}}\tau.
\end{equation}
We suppose that this probability is less than one, that is, there
is at most one fluorescence cycle per pulse. A random number $0\le\alpha_{a}<1$
is picked, and if $\alpha_{a}<g_{a}$ the atom performs a fluorescence
cycle. In such case, another number $\eta_{a}$ taking randomly one of
two possible values $-\kbar/2$ and $+\kbar/2$ is picked to decide
of the direction of the recoil of the atom %
\footnote{That is, from which of the two arms of the standing wave the absorbed
photon comes.}.
The atom decay back to the ground state produces a new recoil in a 
random direction given by another random number $\eta_{sp}=\pm\kbar/2$. 
We suppose that $T\gg\Gamma^{-1}$ \footnote{Typical experimental values for 
cesium are $\Gamma^{-1}\sim30$ ns and $T\sim20$ $\mu$s. Our hypothesis is 
thus well verified.}, 
so that, with respect to the atom dynamics, the whole fluorescence
cycle can be considered as instantaneous. The effect of the spontaneous
emission during the $n^{th}$ kick is thus to change the momentum
$p_{n}\rightarrow p_{n}+\left(\eta_{a}+\eta_{sp}\right)$. This
change in the momentum displaces the point corresponding to the $p_{n}$
in the phase-space of a quantity that can be as large as $\kbar[2\pi]$,
compared to the width of the map $2\pi$. The effect of spontaneous
emission thus increases as the effective Planck constant $\kbar$ of the system
increases, that is, as the system becomes more ``quantum''. Fortunately,
the classical type of dynamics we are interested in here corresponds
to $\kbar < 1$.

\begin{figure} % Fig. 5
\begin{center}\includegraphics[%
  width=8cm]{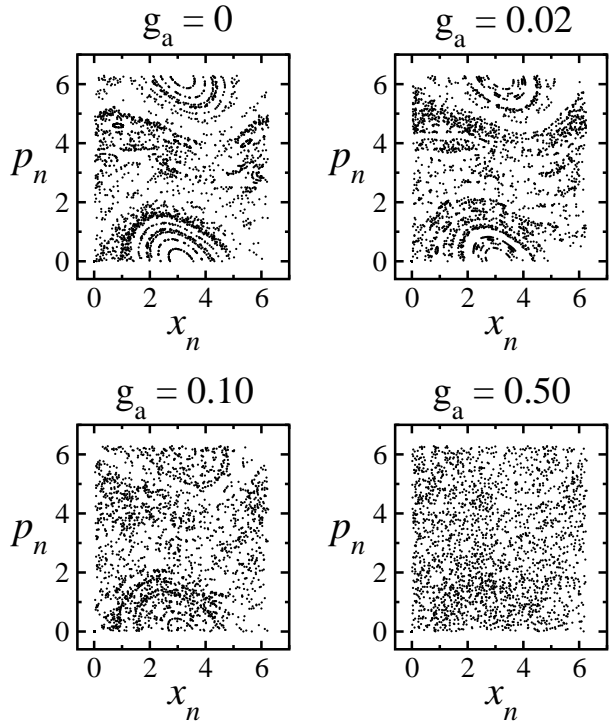}\end{center}
\caption{
\label{fig:KRwithSpontEmission}Influence of the spontaneous emission
in the phase-space reconstruction. Parameters are $K=0.8$ and $F=10^{6}$.}
\end{figure}

Fig.~\ref{fig:KRwithSpontEmission} compares the phase-space reconstruction
obtained with different values of $g_{a}$. The number of restarts
due to bad choices of position/momentum determinations was 
typically two times larger than in
the absence of spontaneous emission. We can estimate 
the maximum spontaneous emission rate per kick still allowing the
reconstruction of the phase map as $g_{max}\sim 0.05$, which means
that, in order to perform this experiment, the parameters should be
chosen so that the localization time be of the order of 
$t_{L}\sim g_{max}^{-1}\sim20$
kicks. However, 20 kicks are not enough to produce a good phase space
portrait; a better combination of parameters would be, e.g.,
$g\sim0.01$ and $t_{L}\sim100$. This values are still compatible with
current experimental values.

\section{Conclusion}
We have proposed and numerically tested a method allowing reconstruction
of the classical dynamics of the ``atom-optics'' kicked rotor
formed by an atom exposed to pulses of an intracavity standing wave.
We have discussed the relevant parameters of the system and showed
that they can fit with real experimental situations. Our method can
be used to probe interesting phenomena, like the transition from the
classical dynamics to the quantum dynamics as the spontaneous emission
rate becomes smaller than the inverse of the localization time. 

It is rather difficult to guess what will happen when this limit is
approached. In the quantum case, the atom cannot be characterized
by a single point in the phase space. This means that the spreading
of the wavefunction must be considered, and the classical notion of
trajectory is not any more a good one. In particular, the
the cavity transmission will be affected by the spatial spreading,
as it allows the atom to ``probe'' different intensity regions.
One must also consider the decoherence effect of spontaneous emission
that reduces the wavepacket if it happens often enough. An interesting
way to explore such situation is to compare a Wigner-function picture
of the quantum phase-space in presence of decoherence with the classical
map \cite{Davidovich_KRDehPhaseSpace_PRE2004}. If the present method
proves able to experimentally explore the classical/quantum transition,
it would be highly valuable for studies of the 
nontrivial relation between classical and quantum chaos.

\begin{acknowledgments}

Laboratoire de Physique des Lasers, Atomes et Mol{\'e}cules (PhLAM) is
Unit{\'e} Mixte de Recherche UMR 8523 du CNRS et de l'Universit{\'e} des Sciences
et Technologies de Lille. HLDSC wishes to acknowledge partial support
by a French Ministry of Research ``ACI Photonique'' fellowship and
by Brazilian Agency Conselho Nacional de Desenvolvimento Cientifico e
Tecnologico (CNPq). CRC thanks Universit{\'e} des Sciences et Technologies de
Lille for a ``professeur invit{\'e}'' fellowship.
\end{acknowledgments}

%\bibliographystyle{apsrev}
%\bibliography{../../../ArtDataBase,../../../Books}

\end{document}